# Chaotic dynamics of strings around the Bardeen-AdS black hole surrounded by quintessence dark energy

Jiayu Xie, Yaxuan Wang, Bing Tang[*]

*Department of Physics, Jishou University, Jishou 416000, China*

We study the motion of a ring string in the background of the Bardeen-AdS black hole surrounded by the quintessence dark energy. The effects of the magnetic monopole charge $\beta$, the quintessence state parameter $\omega_q$, and the quintessence normalization parameter $a$ on the dynamical behavior of the ring string are respectively analyzed. Our numerical results show that the chaotic behavior of string generally becomes stronger with the increase of the quintessence normalization parameter. In particular, the conditions for the existence of chaos are distinctly diverse for two different quintessence state parameters. Furthermore, it is found that the magnetic charge does not significantly affect the chaotic behavior of the string in a specific range.

## I. INTRODUCTION

In recent years, due to an intrinsic non-linearity of the General Relativity, the dynamic of chaos has already been one of the great focus issues in these relativistic systems. On account of the susceptivity to the initial condition, chaos is a kind of random motion and cannot be predicted in deterministically and nonlinearly dynamic systems, leading in turn to appeal to many researchers spread deep exploration to chaotic phenomena. However, the chaos always emerges when the researchers expanded their work to some complicated space-time backgrounds, for example, exploring the dynamic behavior of the charged particle around one black hole including quasi-topological electromagnetism[1], the test particles around deformed electrically and magnetically charged Reissner-Nordström black holes[2], and a particle among the Rindler space in a 4-Dimension black hole having spherically symmetric construction[3]. These researches reveal that the main reason for the occurrence of chaos is that the equation of motion is variable-inseparable and the corresponding nonlinear dynamical system is not integrable, which is led by those complicated couplings. In addition, although the dynamic of a particle is integrable, it has also been proved that the chaotic phenomena can occur in a certain potential, for example, a massless and chargeless particle with the addition of harmonic perturbation in a very near horizon region[4-7].

[*] Corresponding author.
E-mail addresses: bingtangphy@163.com; bingtangphy@jsu.edu.cn

Until now, the motion of a string in space-time has also attracted the interest of many researchers, which is rapidly becoming a vigorous area of research in string theory. The more complex and more intriguing behaviors of the motion of the string can be exhibited via the inherent outstretched nature of string. Aspects of circular strings dynamics in the Schwarzschild black hole with the foundational background have been profoundly studied[8-10]. As pointed out in Refs.[11-13], the influence of the parameters of one charge Reissner-Nordström black hole or one anti-de Sitter(AdS) Gauss-Bonnet black hole causes that dynamical behaviors of one circular string generate a special transformation. There is potential significance to discuss the conversion from order to chaos and the condition of transformation. Moreover, some researchers also pay close attention to chaos bound and its violation[14-17]. Čubrović has performed a systematic study in AdS black hole geometries with multiple horizons[14]. Gao *et al.* have inquired into influencing mechanism of the angular momentum in the charged Kiselev black hole surrounded by an anisotropic fluid[15].

At a classical level, the motion of one string in miscellaneous curved space-time backgrounds is non-integrable. As already shown in Refs.[18-26], the dynamics of the string behave features of chaotic motion, even on the radially symmetric backgrounds[27-29]. In contrast, the motion of a test particle is integrable in most radially symmetric backgrounds. Understanding the difference of motion between a string and a particle in the cosmological or gravitational backgrounds may give us some enlightening views and help us figure out various problems which are associated with classical gravitational singularities. Another motivation to investigate string dynamics in black holes arises from growing interest in chaos near black holes, which is aroused by the approach on studying more complex quantum chaos through some analysis of the dynamic behavior of those relatively simple black holes. Chaos of one string around a black hole is an important topic of contemporary research in physics. As far as we are aware, the characteristic of the chaos in the vicinity of a black hole is linked to its character. Hence, investigating as many various black holes as possible may be available to help us comprehend those chaos phenomena more clearly. However, in masses of curved time-space contexts, string theory is not capable of being resolved analytically. It is widely advised that such kinds of integrability can be stretched to the field theory via taking advantage of the AdS/CFT theory duality[30].

As is well-known, the current cosmological observations forecast the presence of a special kind of energy that pervades the whole of space, called dark energy, which could be one of the origins of the negative pressure. Modern cosmological observations have revealed that about 68.3% dark energy of the energy density constitutes our current universe, according to the standard model of cosmology[31-33]. As one of the most fitting dark energy candidates, the quintessence has attracted much attention, whose state equation $p_q = \omega_q \rho_q$ is satisfied for $-1 < \omega_q < -1/3$. With the help of the phenomenological approach suggested by Kiselev[34], one can effortlessly analyze a black hole encircled by quintessence directly. The regular Bardeen black hole meets the weak energy condition and is regular everywhere, which was first proposed in Ref.[35]. In previous studies, those physical analyses on the matter and thermodynamical quantities have indicated that the Bardeen model is a decent candidate for understanding astrophysical black holes[36-38], and is also considered as a natural particle accelerator[39]. So far, some successive studies on quintessential Bardeen-AdS black hole have also been carried out[40-43]. Unfortunately, very little attention has been paid to chaotic dynamic of strings in the context of a black hole surrounded with dark energy. We are of interest to know the magnetic monopole charge whether still holds a vital role in the chaotic motion of strings around the Bardeen-AdS black hole surrounded by the quintessence dark energy.

In this work, we focus on the dynamic behavior of string around the Bardeen-AdS black hole encircled by the quintessence dark energy. The aim of the present paper is to present one comprehensive work to study the chaotic behavior in the corresponding phase space by considering the influence of the quintessence state parameter $\omega_q$ and the normalization constant $a$. Particularly, we attempt to find out what effect the quintessence on the dynamics of strings near the quintessential Bardeen-AdS black hole. In addition, the impact of the parameter $\omega_q$ corresponding to the different dark energy backgrounds is essential to the motion of the ring string, which is not fully characterized so far. We shall accomplish this work by calculating chaos indicators: Poincaré sections and the maximum Lyapunov exponents.

We organize the present article as follows. In Sect.2, we introduce the solution for one Bardeen-AdS black hole encircled by the quintessence dark energy. And then, the equations of motion of the ring string around the quintessential Bardeen-AdS black holes are derived in Sect.3. The numerical methods and chaos indicators used in this paper are shown in Sect.4. What is more, the solution for the present dynamical system and the characteristics of the dynamic behavior of strings are explored. The last section is used to summarize and discuss the present work.

## II. THE BARDEEN-ADS BLACK HOLE SURROUNDED BY QUINTESSENCE

The Bardeen black hole is one regular spacetime solution of Einstein field equation with the coupling of a nonlinear electromagnetic field. The four-dimensional asymptotic AdS spacetime action can be expressed as[35]

$$S = \int d^4 x \sqrt{-g} [\frac{1}{16\pi} R + \frac{1}{8\pi} \frac{3}{l^2} - \frac{1}{4\pi} \mathcal{L}(\mathcal{F})], \qquad (1)$$

where $R$ and $l$ are the Einstein scalar and the AdS radius, respectively. The first term in brackets signifies the Einstein gravity, the second represents the negative cosmological constant $\Lambda = -3/l^2$, and the third term $\mathcal{L}(\mathcal{F})$ is a nonlinear electrodynamics source Lagrangian, which reads

$$\mathcal{L}(\mathcal{F}) = \frac{3M}{\beta^3} \left( \frac{\sqrt{4\beta^2 \mathcal{F}}}{1 + \sqrt{4\beta^2 \mathcal{F}}} \right)^{5/2}. \tag{2}$$

Here, $\mathcal{F}$ is denoted as $\mathcal{F} = \mathcal{F}_{\mu\nu}\mathcal{F}^{\mu\nu}$, where $\mathcal{F}_{\mu\nu}$ represents the electromagnetic field intensity and $\beta$ corresponds to the positive magnetic monopole charge related to it. In the weak field limit $\mathcal{F} \to 0$, the Lagrangian of the density of the nonlinear electrodynamics is degraded to $\mathcal{L} \sim \beta^{-1/2}\mathcal{F}^{5/4}$, which is marginally strong compared to Maxwell field.

The investigation for the effect of the quintessence on the dynamic behavior of a black hole is inescapable, since the quintessence exists across our entire universe. Taking into account the quintessence, the above mentioned action adopts the following form

$$S = \int d^4x \sqrt{-g} \left[ \frac{1}{16\pi} R + \frac{1}{8\pi} \frac{3}{l^2} - \frac{1}{4\pi} \mathcal{L}(\mathcal{F}) + \mathcal{L}_Q \right]. \tag{3}$$

Altering the action Eq. (3) for the electromagnetic field $\mathcal{F}$ and the metric tensor $g_{\mu\nu}$, respectively, one can write down the equations of motion, namely,

$$G_{\mu\nu} - \frac{3}{l^2} g_{\mu\nu} = 2\left( \frac{\partial \mathcal{L}(\mathcal{F})}{\partial \mathcal{F}} \mathcal{F}_{\mu\nu}\mathcal{F}_\nu^\rho - g_{\mu\nu}\mathcal{L}(\mathcal{F}) \right) + T_{\mu\nu},$$

$$0 = \nabla_\mu \left( \frac{\partial \mathcal{L}(\mathcal{F})}{\partial \mathcal{F}} \mathcal{F}^{\mu\nu} \right). \tag{4}$$

Under this theoretical model, the quintessence is derived from one fluid possessing the energy-momentum tensor $T_{\mu\nu}$. Through the use of the Kislev phenomenological model, the regular-Bardeen black hole surrounded by the quintessence can be constructed as[34]

$$T_t^t = T_r^r = \rho_q,$$

$$T_\theta^\theta = T_\varphi^\varphi = -\frac{1}{2}\rho_q(3\omega_q + 1), \tag{5}$$

where $\rho_q = -3a\omega_q/[2r^{3(\omega_q+1)}]$ denotes the energy density of the quintessence. $\omega_q$ is the quintessence state parameter that takes values in the range $-1 < \omega_q < -1/3$ and $a$ is a positive normalization constant.

By solving Eq. (4), one can get a static spherically symmetry solution, which is

$$ds^2 = -f(r)dt^2 + \frac{dr^2}{f(r)} + r^2(d\theta^2 + \sin^2\theta d\varphi^2), \qquad (6)$$

where $M$ is black hole mass. The $f(r) = 1 - 2Mr^2/(r^2+\beta^2)^{3/2} + r^2/l^2 - a/r^{3\omega_q+1}$ corresponds to the Bardeen-AdS black hole encircled by the quintessence. For one regular Bardeen-AdS black hole in $AdS_4$ spacetime, the corresponding metric function is expressed as $f(r) = 1 - 2Mr^2/(r^2+\beta^2)^{3/2} + r^2/l^2$.

Note that, in the case of without quintessence, viz., $a = 0$, the above space-time is restored to one regular Bardeen-AdS black hole. And one Schwarzschild-AdS black hole encircled by quintessence is derived by reduction for $\beta = 0$. If it has been set as $a = 0$, $\beta = 0$, and $\Lambda = 0$, then we can get the Schwarzschild black hole. The Bardeen black hole, as mentioned in the introduction, is regular everywhere. In the current work, the dynamics of the string near one regular Bardeen-AdS black hole shall be further investigated.

Moreover, the Bardeen-AdS black hole also possesses multiple horizon constructions: two horizons including the event horizon, one event horizon, and no horizon, which depends on those related parameters. On the event horizon $r_h$, the condition $f(r_h) = 0$ should be satisfied, by which one can confirm the parameter $M$. The black hole mass $M$ can be written in terms of the event horizon $r_h$ as

$$M = (1 + \frac{r_h}{l^2} - \frac{a}{r_h^{3\omega+1}}) \frac{(\beta^2 + r_h^2)^{3/2}}{2r_h^2}. \qquad (7)$$

In Fig. 1, the evolution of the mass $M$ concerning to the horizon radius for the various parameters $\beta$, $a$ and $\omega_q$ is plotted. It is clearly seen that the minimum value of the construction Bardeen-AdS black hole mass can be influenced by varying these parameters. In Fig. 1(a), we show that the curve from top to bottom corresponds to a decrease of the value of the magnetic monopole charge $\beta$ in turn when the dark energy parameters are fixed. Thus, the mass extremum of the present black hole increases as the value of $\beta$ increase. From Fig. 1(b), the effect of the normalization factor $a$ on the minimum of the mass $M$ can be observed. Obviously, the increase of $a$ widens the value range of $M$ by causing diminution of the minimum. By comparing Fig. 1(c) with Fig. 1(d), one can realize that the curve turns over for $\omega_q = -2/3$, which indicates that the extreme value of the mass of the quintessential Bardeen-AdS black hole will disappear when the value of $a$ is large enough. Furthermore, no matter what value of $\omega_q$ is chosen, the mass $M$ is same at $r_h = 1$.

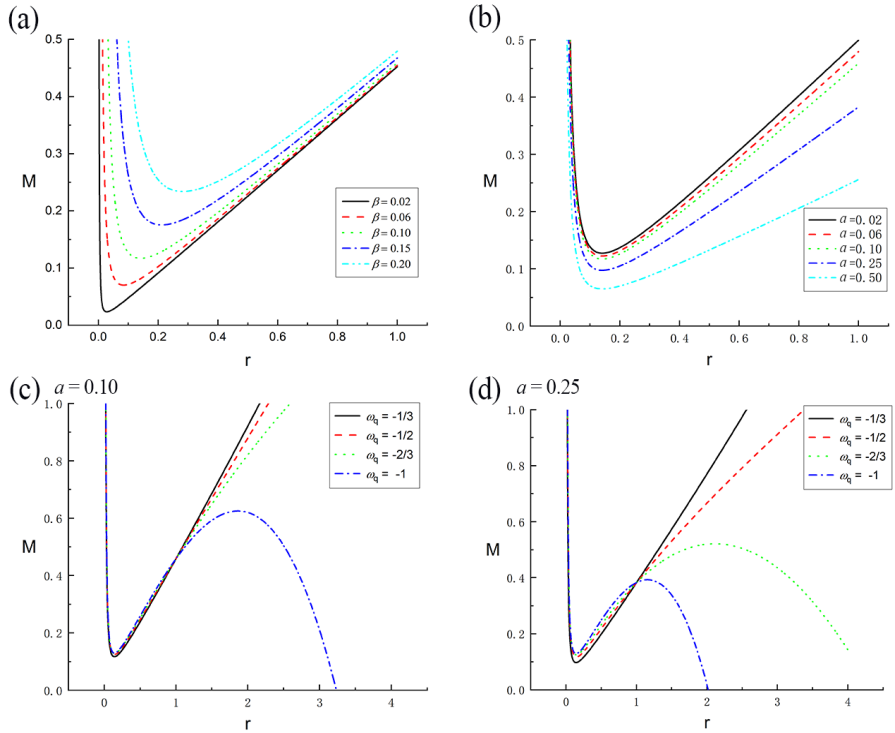

FIG. 1. Variation of mass with respect to the horizon for a set of values of parameters $\beta$, $a$ and $\omega_q$, each parameters are fixed as (a) $a = 0.1$, $\omega_q = -1/3$; (b) $\beta = 0.1$, $\omega_q = -1/3$; (c) $a = 0.1$, $\beta = 0.1$; (d) $a = 0.25$, $\beta = 0.1$.

Based on the above analysis, we can deduce that the mass of the quintessential Bardeen black hole has a minimal value for various values of parameters $a$ and $\beta$. We note that the quintessential Bardeen black hole becomes the state of extreme black hole when the mass of the black hole cannot be reduced any further. What is more, the range of the system parameters are also determined by the mass $M$. In Table. I, we summarize the minimum value of the mass of the quintessential Bardeen-AdS black hole for the results in Fig. 1. According to this data, we can deduce that when the mass of the present black hole is taken to be $M = 0.2$, the range of variation of the magnetic monopole charge needs to satisfy $\beta < 0.20$.

Table I. The minimum of black hole mass $M$ for a set of values of parameters $\beta$, $a$.

| $\beta$ | 0.02 | 0.06 | 0.10 | 0.15 | 0.20 |
|---|---|---|---|---|---|
| $M$ | 0.0233844 | 0.0701507 | 0.116926 | 0.175409 | 0.233919 |
| $a$ | 0.02 | 0.06 | 0.10 | 0.25 | 0.50 |
| $M$ | 0.127318 | 0.122122 | 0.116926 | 0.0974399 | 0.0649638 |

## III. RING STRING IN THE BARDEEN-AdS BLACK HOLE BACKGROUND

Now, let us take into account the motion of one ring string traveling around the Bardeen-AdS black hole surrounded by quintessence. The dynamics of the circular ring string is described in terms of the Polyakov action,

$$S = \frac{-1}{2\pi\alpha'} \int \sqrt{-h} h^{ab} G_{\mu\nu} \partial_a X^\mu \partial_b X^\nu d\tau d\sigma, \qquad (8)$$

where $h^{ab}$ corresponds to the worldsheet metric, $G_{\mu\nu}$ is called the target space metric, $X^\mu$ is the target space coordinates, $\alpha'$ relates the string length, and the superscripts $(a,b) = 1,2$ represent $(\tau,\sigma)$, which are expressed as the coordinates on the worldsheet of the string.

The main string configuration of our concern (depicted in Fig. 2) is described by inserting the following into the aim space:

$$t = t(\tau), \quad r = r(\tau), \quad \theta = \theta(\tau), \quad \varphi = n\sigma. \qquad (9)$$

The ring string has a winding number $n$, which depicts the difference from the particle. Inserting the above ansatz into Eq. (8) in the conformal gauge $h^{ab} = \eta^{ab}$, the Polyakov Lagrangian is explicitly calculated as

$$L = -\frac{1}{2\pi\alpha'}(f\dot{t}^2 - \frac{\dot{r}^2}{f} - r^2\dot{\theta}^2 + r^2 \sin^2\theta n^2). \qquad (10)$$

The Lagrangian in Eq. (10) can generate equations of motion for the cyclic string, which possess the following form

$$\ddot{r} = -\frac{f'}{2f}\dot{r}^2 - \frac{ff'}{2}\dot{t}^2 + rf(\dot{\theta}^2 - n^2\sin^2\theta),$$

$$\ddot{\theta} = -2\frac{\dot{r}\dot{\theta}}{r} - n^2\sin\theta\cos\theta, \quad \dot{t} = \frac{E_n}{f}. \tag{11}$$

Here, the dot in the superscript signifies the derivative with respect to $\tau$ (same below), and the superscript symbol ′ is used hereafter to stand for the derivative with respect to $r$.

Noting that the conformal gauge $h^{ab} = \eta^{ab}$ can result the following nontrivial constraint

$$G_{\mu\nu}(\partial_0 X^\mu \partial_0 X^\nu + \partial_1 X^\mu \partial_1 X^\nu) = 0. \tag{12}$$

What is more, in the background described via Eq. (6) with Eq. (11), a more implicit form on the constraint can be written as

$$\dot{r}^2 + r^2 f \dot{\theta}^2 + n^2 r^2 f \sin^2\theta = E_n^2. \tag{13}$$

The canonical momenta have the following form

$$P_t = -\frac{1}{\pi\alpha'} f\dot{t}, \quad P_r = \frac{1}{\pi\alpha'}\frac{\dot{r}}{f}, \quad P_\theta = \frac{1}{\pi\alpha'} r^2 \dot{\theta}. \tag{14}$$

Employing canonical transform, one can get the according Hamiltonian, which is

$$H = \frac{\pi\alpha'}{2}[fP_r^2 + \frac{P_\theta^2}{r^2} - \frac{P_t^2}{f}] + \frac{1}{2\pi\alpha'} r^2 n^2 \sin^2\theta. \tag{15}$$

Then, by the use of the Poisson bracket, it is not difficult to derive the following canonical equations of motion

$$\dot{t} = -\frac{\pi\alpha' P_t}{f}, \quad \dot{P}_t = 0, \tag{16}$$

$$\dot{r} = \pi\alpha' f P_r, \quad \dot{P}_r = -\frac{\pi\alpha'}{2} f' P_r^2 - \frac{\pi\alpha' P_t^2 f'}{2f^2} + \frac{\pi\alpha' P_\theta^2}{r^3} - \frac{n^2 r \sin^2\theta}{\pi\alpha'}, \tag{17}$$

$$\dot{\theta} = \frac{\pi\alpha' P_\theta}{r^2}, \quad \dot{P}_\theta = -\frac{n^2 r^2 \sin\theta\cos\theta}{\pi\alpha'}. \tag{18}$$

From Eq. (16), one can obtain a relation $P_t = E_n$, which is a constant of motion to characterize the energy. The Hamiltonian constraint, i.e., $H = 0$, results from an equation $T_{ab} = 0$ by fixing the conformal gauge. The constraint given in Eq. (15) stands for the condition of energy conservation in a circular string motion near a Bardeen-AdS black hole surrounded by the quintessence dark energy.

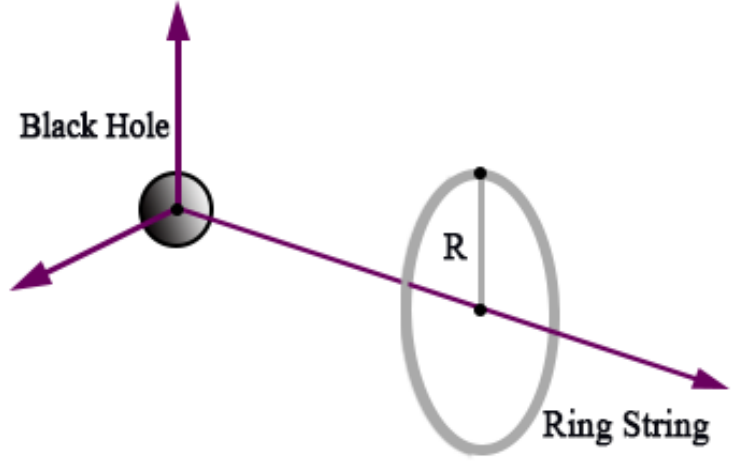

FIG. 2.  Ring string in the Bardeen-AdS black hole background.

Looking back at Fig. 2, it should be noted that in this structure the string radius varies over time as $R(\tau) = r\sin\theta$. The ring string has varied possible modes of motion, which are depicted via the solutions $r(\tau)$ and $\theta(\tau)$. In addition, reviewing the results in Refs.[8,11], we realize that the string radius $R(\tau)$ also helps us to ascertain the various modes. For instance, one possible mode of motion is that the ring string can draw near the present black hole and get dispersive, cross over it. After that, it can be pulled back due to the influence of the gravitational force of the black hole. Under the present mode, both variables $r(\tau)$ and $\theta(\tau)$ oscillate in one approach that the radius $R(\tau)$ is all along greater than the Bardeen-AdS black hole. Thus, one can prospect three modes of motion: 1) the ring string keeps up forever traversing back and forward through the present black hole, 2) the ring string accomplishes lots of such oscillations before falling into the present black hole, 3) the ring string accomplishes lots of such oscillations before flighting to infinity. We are now going to start one numerical research of the present dynamical system.

## IV. NUMERICAL STUDY OF THE DYNAMICAL SYSTEM

Here, the numerical approach for solving the equations of motion of one circular string in the context of Bardeen-AdS black holes is introduced. A reliable numerical method is regarded as a fundamental tool to study nonlinear dynamics, since numerical errors may generate pseudo-chaotic behavior. To secure robust control over error propagation and precision, we shall adopt a seventh-eight order continuous Runge-Kutta method. This method can give well control over the output precision due to its adaptive scheme[44]. This precision can be evaluated at any time via assessing the constraint Eq. (15). The behavior of the corresponding solutions will be typically considered so as to identify chaos in this system, which means the need of adopting the small value of the error tolerance $\delta$.

These numerical calculations and the corresponding results displayed in our figures in the current paper are got by setting, in system Eqs. (16-18), $M=0.2$, $n=1$, $\alpha'=1/\pi$, $E=12$, and $l=15$. By setting the mass of the present black hole near the minimum, it is considered to be an extreme black hole state. Fernando[45] has calculated the null geodesics structure around one Schwarzschild black hole enclosed by the quintessence for $\omega_q=-2/3$. In the current research, we are mainly interested in the effect of the magnetic monopole charge $\beta$ and the normalization constant $a$ on the chaotic dynamics. Additionally, we compare the two cases: $\omega_q=-1/3$ and $\omega_q=-2/3$, and try to find out some interesting results.

### A. Chaos indicators

Valid chaos indicators are significant to analyze the evolutionary process of those chaotic dynamic systems. In spite of the temporal evolution can present one visualized picture on dynamics of one circular string, it is not worthwhile to carry forward. This is due to the fact that the result obtained is not sufficient reliable for the long-time evolution. Hence, for purpose of obtaining a better viewpoint of the circular string dynamics, one needs to consider some chaos indicators, like Poincare sections and the Lyapunov exponents.

The 4D phase space $(r, p_r, \theta, p_\theta)$ for the circular string in the Bardeen-AdS black hole background can be simplified as one 3D space by the use of the energy constraint Eq.(15); specifically we fix $H = 0$. This 3D manifold can be described as a family of 2D concentric tori lying within each other in one normal 3D space. For one integrable nonlinear system, any phase curve possessing any selected initial condition shall turbidly fill one 2D torus of the 3D energy manifold. Thus, one 2D plane in the 3D energy manifold intersecting a family of 2D tori in the transverse orientation can compose one codimension-2 Poincaré section. Perhaps, the Poincaré section is also direct-viewing as a family of the concentric circle in the 3D energy manifold. For one integrable nonlinear system, arbitrary phase curve beginning with one point (lying on one specific circle) on the plane (Poincaré section) shall ultimately cross it again after traversing the torus issuing in one new point (on the same circle), on the same 2D plane. This is referred to as the Poincaré map, which is one highly effective method for disposing of those cumbersome calculations of the varying equations. If the trajectory performs quasi-periodic move on the surface of one torus, known as one Kolmogorov-Arnold-Moser(KAM) torus[46], and the "disintegrating" of varieties of the KAM tori into the chaotic trajectories. For one certain Hamiltonian system, the appearance of chaos may be viewed as the disruption of the circle in the Poincaré map. That is to say, the more dispersed the Poincaré section for the Hamiltonian system, the more chaotic it will.

One of significant features of chaos is the susceptive reliance on the initial condition, which signifies that for arbitrary point $r(\tau)$ in the four-dimensional phase space, thus there exist at least a point optionally close to $r(\tau)$ that lapses from it. Let us take into account two points in the four-dimensional phase space, whose distance are denoted by $\Delta x(x_0, 0)$. The separation between these two points is also one function of the original position, which is written as $\Delta x(x_0, \tau)$. The Lyapunov exponent is one significant index that depicts the degree of separation of these infinitesimally near trajectories. Theoretically, the Lyapunov exponent can be defined as

$$\lambda = \lim_{\tau \to \infty} \frac{1}{\tau} \ln \frac{\|\Delta x(x_0, \tau)\|}{\|\Delta x(x_0, 0)\|}. \qquad (19)$$

In the present work, we shall apply an algorithm proposed by Wolf[47], which is identified as the most 'robust' program, peculiarly in specific strongly chaotic circumstances. In the case of $\lambda > 0$, the orbit is chaotic. However, if $\lambda = 0$, then the corresponding orbit is viewed as order. In principle, it is more efficient to apply log-log curve to depict the dynamic behavior. Under this agreement, the motion is identified as ordered when $\log_{10}|\lambda|$ decreases linearly as $\log_{10}(\tau)$ increases. If $\log_{10}|\lambda|$ exponentially varies as $\log_{10}(\tau)$ increases, nevertheless, then the motion is regarded as chaotic. Lyapunov indicators have been widely applied in those higher-dimensional spaces. In particular, they can be devoted to examine the precision of the Poincaré section approach. In our research, we shall calculate numerically the Poincaré sections and the maximum Lyapunov exponents to investigate the chaotic dynamics of the present system.

### B. Strings dynamics around the regular Bardeen-AdS black hole

If the quintessence is absent, then there is no singularity at the origin. In this case, the Bardeen-AdS black hole is referred to as regular black hole or the nonsingular black hole, which possesses an event horizon. Here, we descript the dynamics of one ring string in the context of regular Bardeen-AdS black hole. In Fig. 3, we show the numerical solution of $r(\tau)$ for the magnetic monopole charge $\beta = 0.02$ and $\beta = 0.15$. Without loss of generality, let us consider the initial conditions: $r(0) = 10$, $p_r(0) = 2.670855$, $\theta(0) = 0$, and $p_\theta(0)$ given by the constraint Eq.(15). Although they look regular, it is hard to see any specific pattern behind the amplitude and frequency of such oscillations.

Comparing Fig. 3(a) with Fig. 3(b), we do not find any significant difference between the two cases. In order to further explore the effect of magnetic monopole charge $\beta$ on string dynamics, we consider the string dynamics as a function of $\beta$, particularly for the magnitude of the charge parameter close to the limit. Here, the

Poincaré sections are used to further observe the effect of changing the magnetic monopole charge $\beta$. In Fig. 4, we display the Poincaré sections at $\theta = 0$ for different values of the magnetic monopole charge $\beta$. These plots with different shapes and colors of points describe different values of the magnetic monopole, respectively.

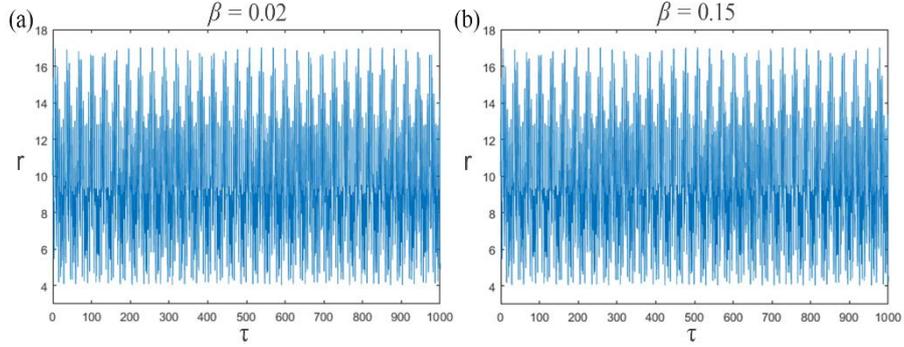

FIG. 3. The solution $r(\tau)$ for the circular string in the regular Bardeen-AdS black hole background for increasing values of the magnetic monopole charge (a) $\beta = 0.02$ and (b) $\beta = 0.15$.

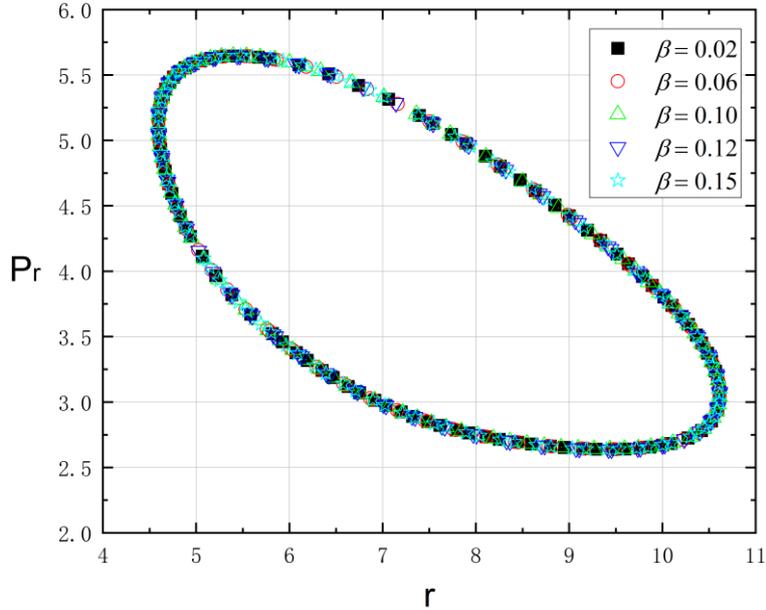

FIG. 4. Plots depict the Poincaré sections at $\theta(\tau) = 0$ for the solutions $r(\tau)$ with increasing values of the magnetic monopole charge $\beta$ of the regular Bardeen-AdS black hole.

In Fig. 4, one can clearly see that the Poincaré sections displays the regular KAM tori for various values of $\beta$. However, we note that the area bounded by the Poincaré sections keeps almost unchanged as we increase the magnetic monopole charge of the circular string for the fixed initial condition, which means that the volume of the phase space remains nearly constant. Hence, we can deduce that the dynamics of the string seems to be insensitive to the variations of the magnetic monopole charge within the desirable range of the magnetic monopole charge. Furthermore, due to the presence of horizon we find that there exist some upper bounds on the magnetic monopole charge. According to our calculations, further increasing the value of the magnetic monopole charge will lead to the unbounded situations. Thus, we were unable to explore the effect of larger values of $\beta$ on string dynamics.

### C. Strings dynamics around the quintessential Bardeen-AdS black hole

Now, let us turn to focus on the chaotic dynamics of strings near the Bardeen-AdS black hole surrounded by quintessence, which includes quintessence parameters $a$ and $\omega_q$. Our aim is to analyze the influence of the normalization constant on dynamics of string for the two quintessence cases. The parameters $a$ and $\omega_q$ are related to quintessence energy density. Under certain initial conditions, we have showed that the string is moving quasi-periodically in the context of AdS space without the quintessence. Here, the same initial conditions shall be chosen to study numerically the influence of the quintessence on the dynamic behavior of strings. In Fig.5, we show the solution numerical of $r(\tau)$ in two cases of the quintessence state parameter, i.e., $\omega_q = -1/3$ and $\omega_q = -2/3$. One can clearly see that the results of the two quintessence state parameter cases are obviously different. Furthermore, compared with the case without quintessence in Fig. 3, the left $r(\tau)$ curve for the case of $\omega_q = -1/3$ is more complex and chaotic, while the right $r(\tau)$ curve for the case of $\omega_q = -2/3$ is more regular. This result indicates that the choice of different quintessence state parameter $\omega_q$ may affect the dynamical behavior of the string. Although Fig. 5(b) seems to regular, it is very hard to find any unambiguous pattern behind the amplitude and frequency of these oscillations. Additionally, Fig. 6 displays the phase curve in the $(r, p_r)$ plane corresponding to those solution solutions $r(\tau)$ given in Fig. 5. The phase curve seems to be densely filled in a given region of the $(r, p_r)$ plane, which implies possible existence of one extremely complicated pattern characteristic of deterministic chaotic systems. Fig. 5(b) appears to show three or four repeating structures at a certain time intervals, which may give the wrong impression of periodic orbit appearance. In Fig. 6(b), we note that there exists one segmented structure in the phase space, which agrees with this inference. Furthermore, it will be further clarified by plotting the corresponding Poincaré sections.

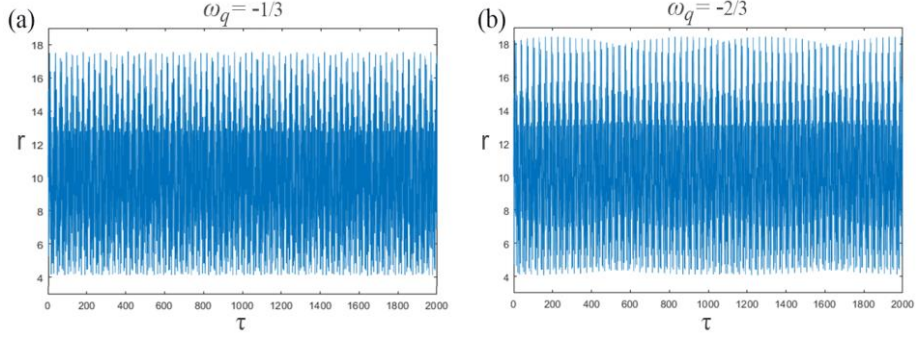

FIG. 5. The solutions $r(\tau)$ for the ring string in the quintessential Bardeen-AdS black hole with different values of the quintessence state parameter (a) $\omega_q = -1/3$ and (b) $\omega_q = -2/3$, by setting $\beta = 0.02$, $a = 0.02$, $r_0 = 10$, $p_r(0) = 2.670855$ and $\theta(0) = 0$. Here, initial conditions used are the same as those in Sect. B.

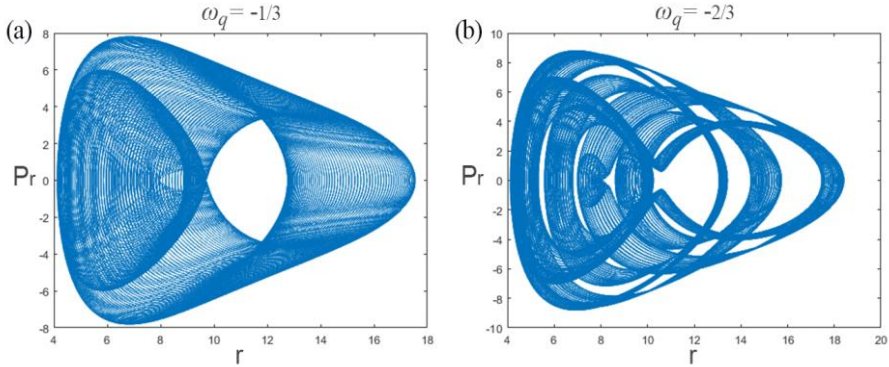

FIG. 6. Phase curves corresponding to those solutions presented in Fig. 5, projected into the $(r, p_r)$ plane.

In Figs. 7 and 8, we plot the Poincaré sections in the plane ($\theta = 0$, $p_\theta > 0$) for the case of $\omega_q = -1/3$, where the ranges of the normalization constant are $a = 0.009 - 0.08$ and $a = 0.08 - 0.11$, respectively. With the increase of $a$, especially up to $a = 0.08$, the points of the Poincaré section appears to change in a complex way. Therefore, in order to better observe its changes, the interval

$a = 0.08 - 0.11$ is represented in separate figures. Because the Poincaré section requires a long enough run to reach a positive conclusion, we consider trajectories within $\tau < 2000$. In Fig. 7, we find that the area enclosed by the KAM tori shown in the Poincaré section becomes larger when we increase the normalization constant of the system. At $a = 0.08$, we notice that there are some slight destructions in the tori, which is one sign of the appearance of chaos. For a given Hamiltonian system, if the motion is viewed as nonchaotic, then the plotted points take shape one closed curve in the two-dimensional phase plane since one regular orbit can move on one torus in the phase space and the curve is one cross section of the tori. Theoretically, the emergence of chaotic motion may be viewed as the disruption of the circle in the corresponding Poincaré section. In the interval of $a = 0.009 - 0.08$, as $a$ increases, the whole process of variation in tori is expansion of area so that fracture of coherent curve and eventually the tori is broken. This result shows that the system becomes chaotic with the increase of the normalization constant for a fixed value of magnetic monopole charge and other initial conditions.

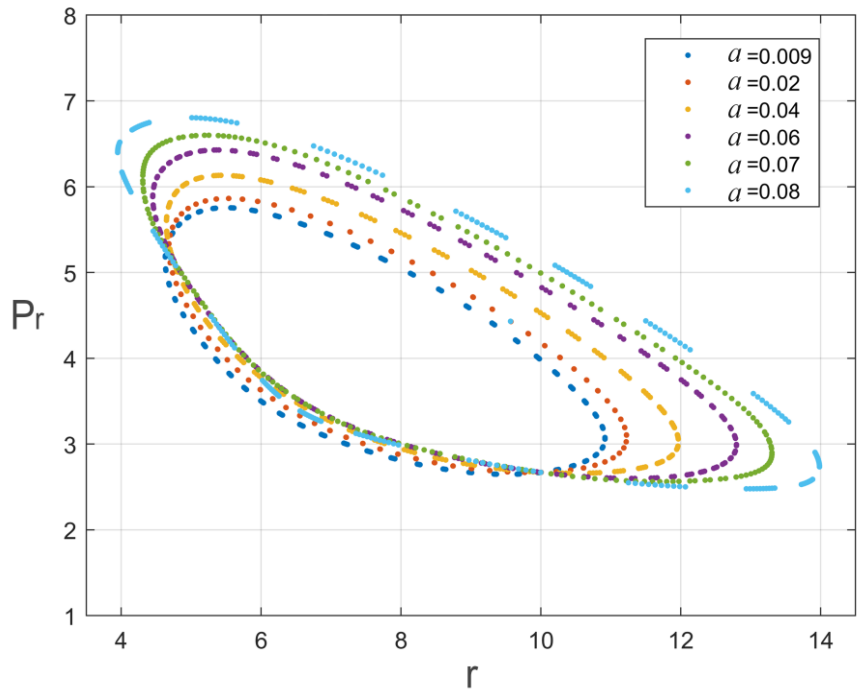

FIG. 7. The Poincaré sections on the plane $\theta(\tau) = 0$, $p_\theta(\tau) > 0$ with increasing values of the normalization constant $a = 0.009 - 0.08$ for $\omega_q = -1/3$ and $\beta = 0.02$.

However, at $a > 0.08$, the variation of the KAM tori becomes very complicated with increasing the normalization constant $a$. As shown in Fig. 8(a)-(b), one torus begins to break up into two. In Fig. 8(b), one can clearly observe that the torus is gradually dissolved by increasing the normalization constant to $a = 0.09$. As $a$ continues to increase, the scattered tori strangely decrease and tend to neatly clustered, which is displayed in Fig. 8(b)-(d). Especially, when $a > 0.1$, we note that the Poincaré section reverts to one torus, and the area bounded by the Poincaré sections also increases as $a$ increases, as presented in Fig. 8(e)-(f). In the phase space, the simultaneous presence of the tori and points that are dispersed from a certain part of the tori can be clearly detected through the Poincaré section, which means that the system is weakly chaotic. When the quintessence normalization constant $a = 0.11$, such structures can be observed in the corresponding Poincaré section, which indicates that the dynamics behavior of the classical circular string in the quintessential Bardeen-AdS black hole background should be weakly chaotic. In addition, our calculations indicate that the string gets trapped by black hole in a short time when $a > 0.11$.

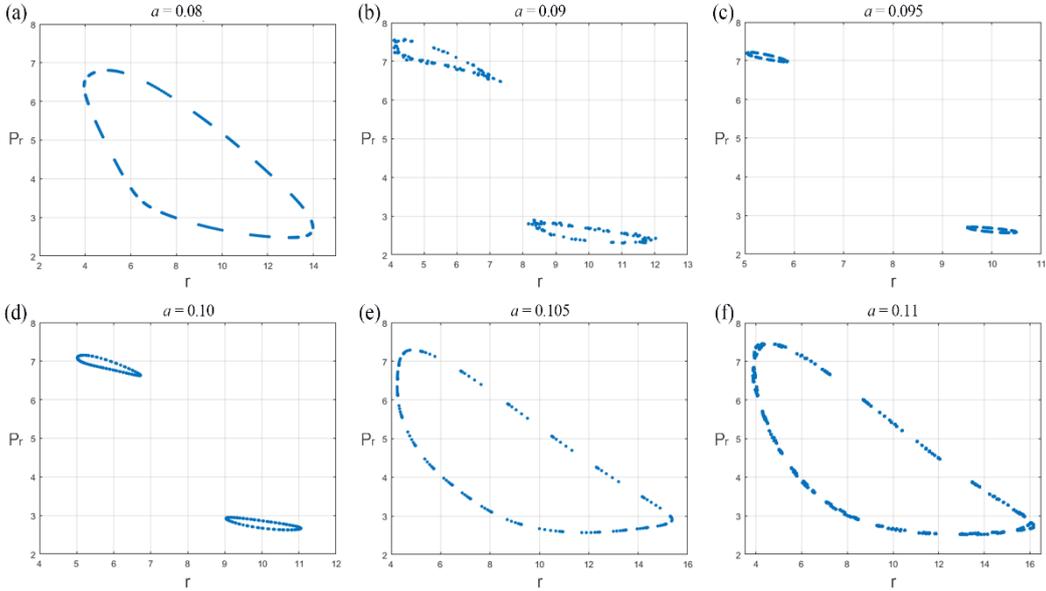

FIG. 8. The Poincaré sections at $\theta(\tau) = 0$, $p_\theta(\tau) > 0$ with increasing values of the normalization constant $a = 0.08 - 0.11$ for $\beta = 0.02$ and $\omega_q = -1/3$.

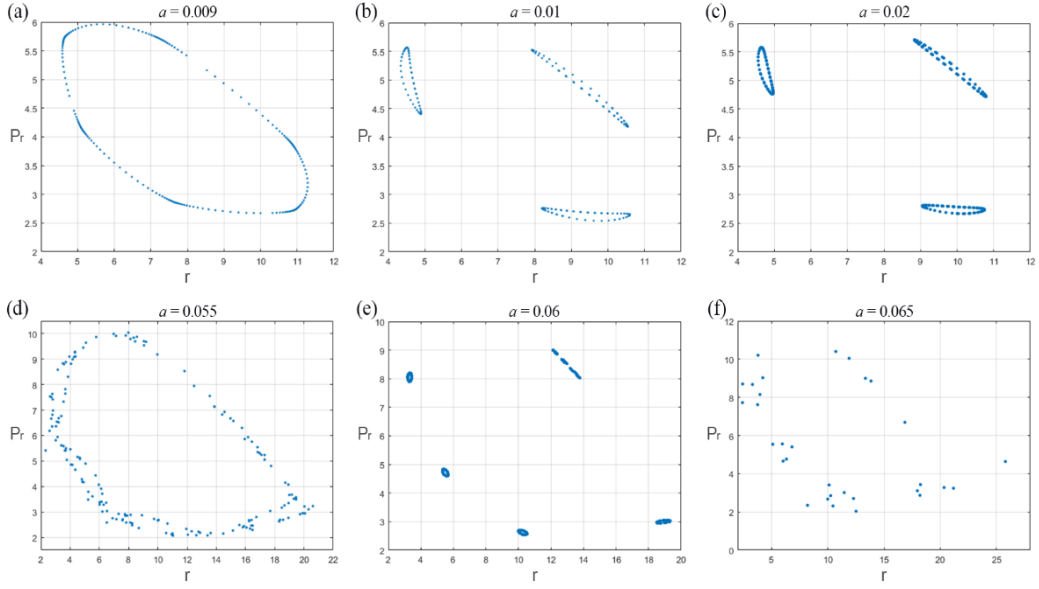

FIG. 9. The Poincaré sections for the solutions $r(\tau)$ with increasing values of the normalization constant $a = 0.009 - 0.065$ for $\beta = 0.02$ and $\omega_q = -2/3$.

To compare to another case of quintessence, in Fig. 9, we exhibit the Poincaré sections in the plane ($\theta = 0$, $p_\theta > 0$) corresponding to the case of $\omega_q = -2/3$. Obviously, with the increases of the normalization constant $a$, the behavior of the present system switches from quasiperiodic to chaotic, where the KAM tori are destructed. Especially for the case of $a = 0.065$ in Fig. 9(f), we can clearly see that the tori are entirely composed of discrete points, which suggests that the chaotic behavior becomes stronger as the normalization constant $a$ increases. Different from the case of $\omega_q = -1/3$, here the chaotic nature occurs at relatively lower value of the normalization constant $a$ and its minor change can cause obvious transitions from quasiperiodic regime to chaos. In addition, with the increase of $a$, the small tori also seem to become small and neat. Similarly, outside of the interval $a < 0.065$, we find that the string will be captured in a short time as the normalization constant $a$ continues to increase, which means that one can observe the chaotic motion of string around the black hole when the variation of the normalization constant is in a smaller range. Most surprisingly, our calculations display that the persistence oscillation of the string gets again when the normalization constant is fixed as $a = 0.10$. The Poincaré section of this special value is shown to be a point, which is a sign of the periodic motion. In Fig. 10, we plot the 3D phase space of this exceptional case, and the corresponding maximum Lyapunov exponents is presented in Fig. 11(d).

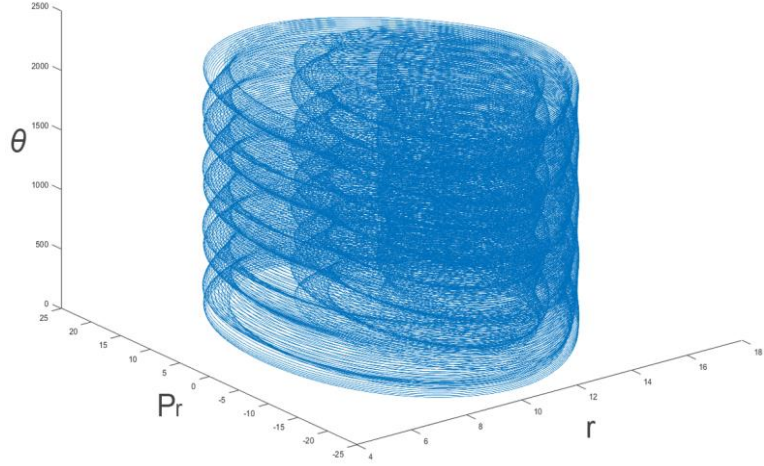

FIG. 10. The phase space $(r, p_r, \theta)$ plots for the ring string in the quintessential Bardeen-AdS black hole background with a specific value of the normalization constant $a = 0.1$, for $\beta = 0.02$ and $\omega_q = -2/3$.

Next, let us study the properties of string dynamics by numerically calculating the maximum Lyapunov exponents. In Fig. 11, we plot the maximum Lyapunov exponents by taking $\log_{10}|\lambda|$ as a function of $\log_{10}(\tau)$ for several modes of asymptotic motion. Fig. 11(a) corresponds to the situation in Fig. 8(f), where the motion of the string is weakly chaotic. In Fig. 11(a), we observe that the orbit motion is sustained oscillation. Fig. 11 (b) corresponds to the situation in Fig. 9(f) and Fig. 11(c) corresponds to the case of beyond the range of the normalization constant in Fig. 9, where both these two curves are perpendicular to the vertical coordinate. This accident is owing to the fact that the distance between the two neighboring orbits becomes extremely short with the increase of time. Obviously, the orbit motions in Fig. 11(b) and 11(c) are collapsed, which means that the string is trapped by the present quintessential Bardeen-AdS black hole. In other word, $\log_{10}|\lambda|$ varies nonlinearly with $\log_{10}(\tau)$. Therefore, both orbit motions in Figs. 11(a) and 11(b) are chaotic, while in Fig. 11(c) the string is extremely quick captured by the present black hole so that there is not enough time to accomplish the stable value of the Lyapunov exponents to detect chaos. In addition, Fig. 11(d) corresponds to the situation in Fig. 10, which is an exceptional case of the motion of strings around the Bardeen-AdS black hole with large value of the normalization constant $a$. In Fig. 11(d), we can see that $\log_{10}|\lambda|$ decreases linearly with $\log_{10}(\tau)$ increasing, which suggests that the orbit in Fig. 11(d) is ordered. As mentioned in the section 3, we have surmised that the motion of the classical ring string in the background of the Bardeen-AdS black hole may contain three asymptotic patterns. Nevertheless, the present Bardeen-AdS black hole is surrounded by quintessence so that our initial conditions yielding escape to infinity form one zero measure set. When the normalization constant $a$ is outside the range, which is shown in Figs. 8 and 9, the motion of the string is unstable and will be captured by the present quintessential Bardeen-AdS black hole after a short run time ($\tau < 100$).

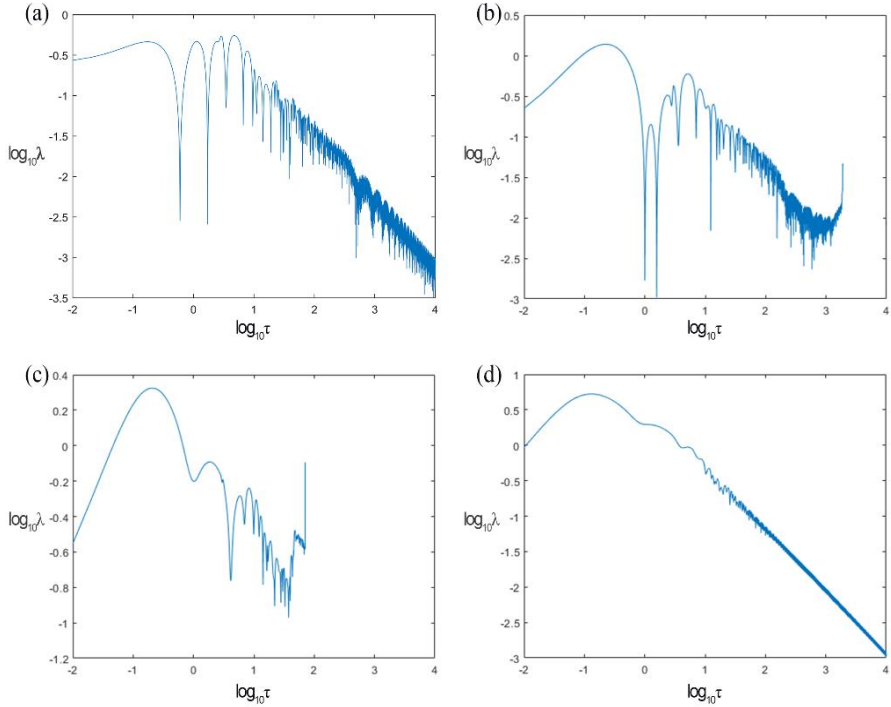

FIG. 11. The maximum Lyapunov exponents of the string trajectory for various asymptotic modes of motion: (a) eternal oscillation; (b) chaotic before capture; (c) capture; (d) order, the parameters setting corresponds to Fig. 8(f), Fig. 9(d), $a = 0.07$ in Fig. 9, and Fig. 10, respectively.

Finally, we further calculate numerically the maximum Lyapunov exponents to analyze the influence of magnetic monopole charge $\beta$ on string dynamics around the Bardeen-AdS black hole surrounded by quintessence. In Fig. 12, we display the maximum Lyapunov exponents for various values of $\beta$, where relevant parameters settings are consistent with the conditions of the occurrence of chaos in Fig. 9. It is found that the string first goes through a long period oscillation and then falls into the quintessential Bardeen-AdS black hole. In these cases, the motion of the ring string is chaotic and finally it is captured by the present quintessential Bardeen-AdS black hole. Thus, we can infer that the choice of the value of $\beta$ has no significant influence on the chaotic dynamics of the string. Furthermore, we note that the captured time of the string doesn't linearly vary with the magnetic charge $\beta$. That is to say, the captured time either increases or decreases as the value of $\beta$ increases. In principle, the decreasing captured time signifies that it is easier for the circular string to be captured by the present quintessential Bardeen-AdS black hole. Thus, there is no certain rule to the influence of $\beta$ on the captured time of the string. Hence, we can deduce that the chaotic dynamics of the circular string around the present quintessential Bardeen-AdS black hole seems to be unresponsive to the magnetic monopole charge, which is analogous to the case of the regular Bardeen-AdS system.

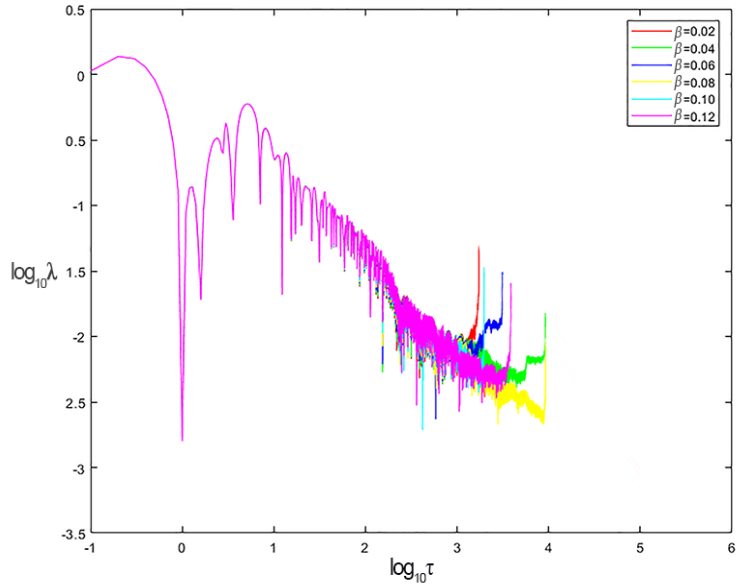

FIG. 12. The maximum Lyapunov exponents of the string trajectory for various magnetic monopole charge $\beta$. Here, we chose $a = 0.055$ and $\omega_q = -2/3$.

# V. CONCLUSION AND DISCUSSION

In the current research, we have studied the dynamics of one classical ring string motion in a quintessential Bardeen-AdS black hole background. The set of nonlinear coupled differential equations governing the string dynamics have been numerically solved. When the normalization constant $a$ is sufficiently large, there exist a cosmological horizon, which is separated by some distance from the event horizon. In this case, there is no minimum limit on the mass of the quintessential Bardeen-AdS black hole. Above all, we have focused on the effect of the magnetic monopole charge $\beta$ and the normalization constant $a$ on the chaotic behaviors of the ring string. Furthermore, the occurrence conditions and properties of chaotic motion of the classical ring have also been analyzed.

In the case of one circular string around the regular Bardeen-AdS black hole background, we have observed that the string trajectories are the quasiperiodic KAM tori, which manifests the dynamic behavior of this system is nonchaotic in our parameter setting. Although, we have not exhibited the change of the Poincaré sections as the energy increases, a similar behavior is to be forecasted as shown in the charge RN-AdS black hole background in which the system becomes more chaotic with the energy of the ring string increasing[11]. Strangely, the analysis of the Poincaré sections has revealed that the dynamics behavior of the ring string is insensitive to the magnetic monopole charge, since the metric function itself depends on $\beta$ sensitively and complicatedly. For being convenient to study, the magnetic monopole charge has been restricted to a small range, which corresponds to the situation of the presence of horizon.

We have also chosen those fixed parameters of the regular Bardeen-AdS black hole background to further study the behavior of the ring string around the Bardeen-AdS black hole surrounded by quintessence changes from nonchaotic to chaotic. We have found that the behavior of the ring string is quite complicated and varied after introducing the quintessence. It has been shown that the normalization constant $a$ performs an instrumental role in the chaotic dynamics of ring string motion. Our results have showed that, in the background of the quintessential Bardeen-AdS black hole, the motion of the ring string is weakly chaotic for the case of $\omega_q = -1/3$ and the stronger chaotic for the case of $\omega_q = -2/3$. Taking into account the quintessence dark energy, one can deduce that its effect emerges very similar to one enhancing/damped mechanism. For example, by studying the thermodynamics of Bardeen-AdS black hole with the quintessence, Ahmed Rizwan *et al*. have found that the intermolecular forces in gases increases slightly as the value of $\omega_q$ decreases[41]. In addition, we have examined the maximum Lyapunov exponents to confirm our findings, and found that the dynamic behavior of the string is also not affected by the magnetic monopole charge in the quintessential Bardeen-AdS black hole background. By contrary, by investigating the thermal chaos in the same background, Wang and Liu have showed that the magnetic monopole charge $\beta$ determine the critical amplitude of temporally periodic perturbation, which is the key value of the emergence of the temporal chaos[43].

In the future research, a great many works and issues deserve further study. First of all, we will study the dual interpretation of the ring string around more black holes surrounded by the quintessence dark energy. Furthermore, we will also extend the present study to some test particles, such as the neutral, the scalar test particle or the test magnetic dipole.

## ACKNOWLEDGMENTS

This theoretical study was supported by the National Natural Science Foundation of China under Grant Nos. 11875126 and 11964011, and the Natural Science Fund Project of Hunan Province under Grant No. 2936.